\documentclass[prl,twocolumn,amssymb,floatfix,amsmath,showpacs,superscriptaddress]{revtex4}
\usepackage{epsfig}

\usepackage{amssymb}
\usepackage{amsmath}
\usepackage{amsfonts}
\usepackage{bbold}
\usepackage{bm}
\usepackage{graphicx}
\usepackage{braket}
\usepackage{color}

\newcommand{\nn}{\nonumber}
\newcommand{\ud}{{\textrm{d}}}
\newcommand{\bk}{{\bf k}}
\newcommand{\br}{{\bf r}}

\newcommand{\llangle}{\langle\kern-.25em\langle}
\newcommand{\rrangle}{\rangle\kern-.25em\rangle}

\begin{document}
\title{Bose and Mott Glass Phases in Dimerized Quantum Antiferromagnets}

\author{S.~J. Thomson}
\affiliation{SUPA, School of Physics and Astronomy, University of St.~Andrews, North Haugh, St. Andrews, KY16 9SS, United Kingdom}
\affiliation{ISIS Facility, Rutherford Appleton Laboratory, Chilton, Didcot, Oxfordshire, OX11 0QX, United Kingdom}
\author{F. Kr\"uger}
\affiliation{ISIS Facility, Rutherford Appleton Laboratory, Chilton, Didcot, Oxfordshire, OX11 0QX, United Kingdom}
\affiliation{London Centre for Nanotechnology, University College London, Gordon St., London, WC1H 0AH, United Kingdom}

\date{\today}

\begin{abstract}
We examine the effects of disorder on dimerized quantum antiferromagnets in a magnetic field, using the mapping to a lattice 
gas of hard-core bosons with finite-range interactions. Combining a strong-coupling expansion, the replica method, and 
a one-loop renormalization group analysis,  we investigate the nature of the glass phases formed. We find that 
away from the tips of the Mott lobes, the transition is from a Mott insulator to a compressible Bose glass, however the compressibility 
at the tips is strongly suppressed. We identify this finding with the presence of a rare Mott glass phase not previously described by any 
analytic theory for this model and demonstrate that the inclusion of replica symmetry breaking is vital to correctly describe the glassy 
phases. This result suggests that the formation of Bose and Mott glass phases is not simply a weak localization phenomenon 
but is indicative of much richer physics. We discuss our results in the context of both ultracold atomic gases and spin-dimer materials.
\end{abstract}


\maketitle


The disordered Bose-Hubbard model is an ideal system for the thorough study of the effects of disorder on strongly 
interacting quantum systems. Ultracold atoms  in optical lattices \cite{Jaksch+98,Greiner+02,Bloch05,Bakr+09,Sherson+10} 
perhaps offer the most direct experimental system in which to realize Bose-Hubbard physics, however the small system sizes and 
destructive nature of many measurements limit the efficacy of experiments. Dimerized quantum antiferromagnets present a 
compelling alternative environment due to an exact mapping to a lattice gas of bosons with hard-core repulsion 
\cite{Matsubara+56,Giamarchi+08,Zapf+14}. These systems consist of lattices of pairs of spins (dimers) which, in 
the ground state, are all in a singlet configuration. This state can be viewed as an `empty' lattice while a local 
triplet excitation can be thought of as a site occupied by a spin-1 boson (`triplon'). 

Condensation of these bosons corresponds to exotic magnetically ordered states seen in materials such as TlCuCl$_3$ 
\cite{Nikuni+00,Ruegg+03,Merchant+14}, ${\mathrm{Cs}}_{2}{\mathrm{CuCl}}_{4}$ 
\cite{Coldea+02,Radu+05}, BaCuSi$_2$O$_6$ \cite{Sebastian+06,Batista+07}, SrCu$_2$(BO$_3$)$_2$ 
\cite{Kodama+02,Jaime+12}, and Ba$_3$Mn$_2$O$_8$ \cite{Samulon+10,Suh+11,Samulon+11}. These systems provide excellent 
experimental setups to probe quantum critical behavior through field- and pressure-tuning, and have motivated some notable 
theoretical works based on bond-operator techniques \cite{Roesch+07,Doretto+12,Joshi+15a,Joshi+15b}. 

Recent experiments on disordered quantum antiferromagnets have seen evidence for novel glassy phases, particularly in 
bromine-doped dichloro-tetrakis-thiourea-nickel (DTN) \cite{Yu+12} where both Bose and Mott glass 
phases of bosonic quasiparticles have been observed. Such phases have also been seen in other materials 
\cite{Hong+10,Stone+11,Yamada+11,Wulf+11} and in quantum Monte Carlo simulations 
\cite{Roscilde+07,Sengupta+07,Yu+10,Ma+14,Wang+15}. 
\begin{figure}[t!]
\begin{center}
\includegraphics[width=  0.6\linewidth]{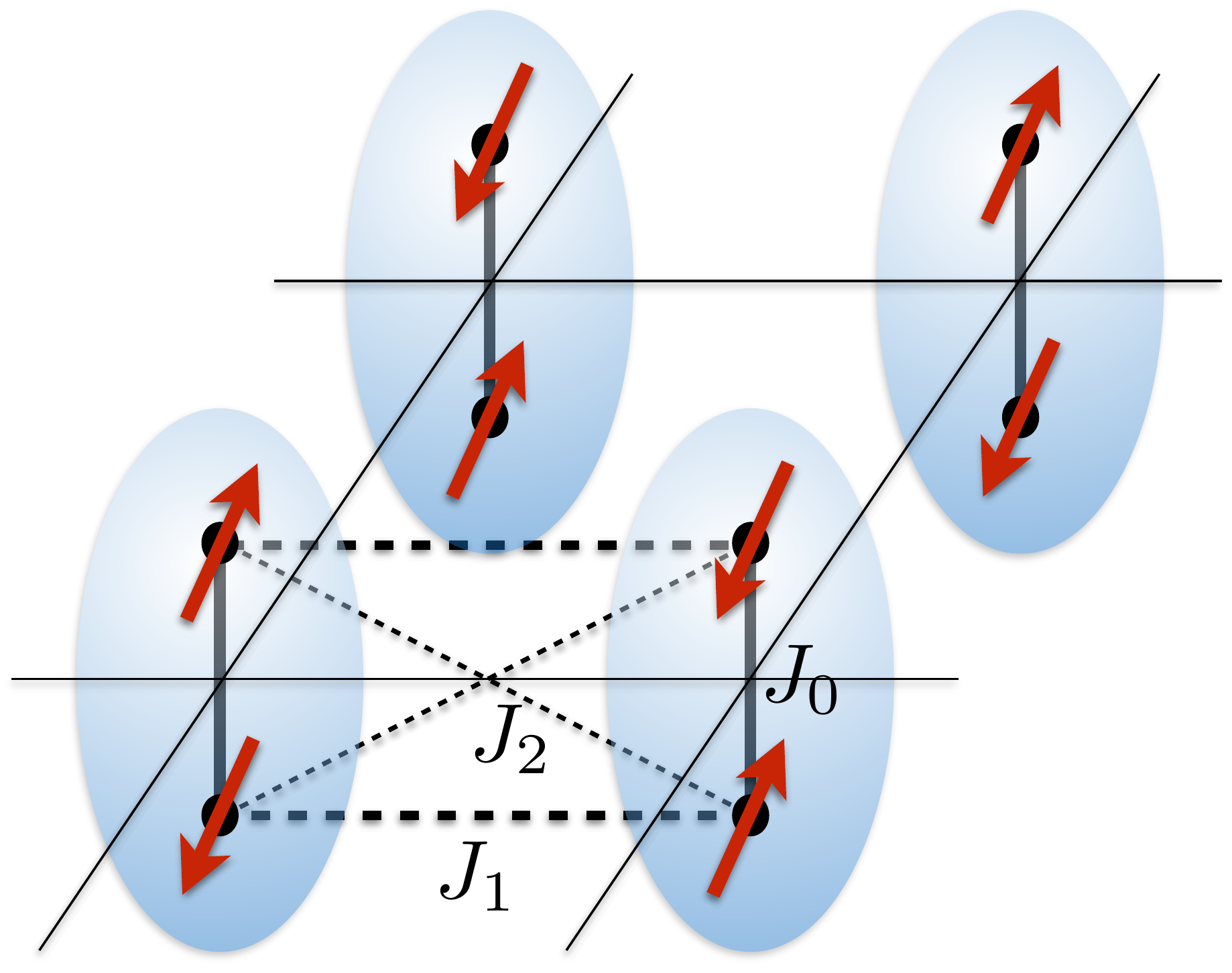}
\caption{(Color online) Square lattice of dimers with intra-dimer coupling $J_0$. Adjacent dimers are coupled by exchanges $J_1$ and $J_2$.}
\label{fig.1}
\end{center}
\end{figure}

Motivated by these experimements, in this Letter we present an analytic treatment of dimerized quantum antiferromagnets with weak 
intra-dimer bond disorder using the hard-core boson formalism. We perform a strong coupling expansion \cite{Freericks+96,Sengupta+05} 
combined with a replica disorder average to derive an effective field theory. From a renormalization group (RG) analysis we obtain the phase 
boundaries between the gapped magnetic states -- or in boson language,  incompressible Mott insulating states --  and the adjacent 
spin-glass phases. We show that away from the tips of the Mott lobes, the spin glass is equivalent to a compressible Bose glass, 
while at the tips we have strong indication for the existence of an incompressible Mott glass.  

The finite compressibility of the Bose glass turns out to be a direct consequence of replica symmetry breaking (RSB), a mathematical 
property signifying the non-ergodic nature of the glassy states. Our work clearly shows that RSB in disordered Bose-Hubbard models is 
directly linked to the physical properties of the glassy phases and that it finds a natural interpretation in terms of analogous disordered spin 
systems.

We start from a Hamiltonian describing a lattice of coupled dimers of $S=1/2$ spins, subject to single-dimer anisotropy $D$ \cite{anisotropy} and 
 magnetic field $h$, 
\begin{eqnarray}
\mathcal{H} &= &\sum_{i} \left[ J_0 \medspace \hat{\bf S}_{i1} \cdot \hat{\bf S}_{i2}-D(\hat{S}_{i1}^z+\hat{S}_{i2}^z)^2 -h(\hat{S}_{i1}^{z}+\hat{S}_{i2}^{z}) \right] \nn\\
& & \quad \quad + \frac12 \sum_{i \neq j}\sum_{m,n} J_{ijmn} \medspace \hat{\bf S}_{im} \cdot \hat{\bf S}_{jn},
\end{eqnarray}
where $i,j$ label the dimers and $m,n=1,2$ the component spins of the dimers. This Hamiltonian is quite generic and describes most of the 
aforementioned dimer compounds \cite{Nikuni+00,Ruegg+03,Merchant+14,Coldea+02,Radu+05,Sebastian+06,Batista+07,Roesch+07,Kodama+02,Jaime+12}. 
For simplicity, we assume that the dimers are located on the sites of a $d$-dimensional hyper-cubic lattice 
and that the couplings between dimers are isotropic along different bond directions. To be specific, we consider 
superexchanges $J_1$ and $J_2$ between adjacent dimers (Fig.~\ref{fig.1}), where $J_0\gg J_1>J_2>0$. 

The mapping of the dimerized quantum antiferromagnet to a model of hard-core bosons is achieved by expressing the spin operators of each dimer
 in terms of singlet and triplet bond operators \cite{Roesch+07,Doretto+12,Joshi+15a,Joshi+15b}, 
 $\hat{s}^\dagger\ket{0}=\frac{1}{\sqrt{2}}(\ket{\uparrow \downarrow}-\ket{\downarrow \uparrow})$,  $\hat{t}_{+}^{\dagger}\ket{0}=\ket{\uparrow \uparrow}$,
 $\hat{t}^\dagger_0\ket{0}=\frac{1}{\sqrt{2}}(\ket{\uparrow \downarrow}+\ket{\downarrow \uparrow})$,  and 
$\hat{t}_{-}^{\dagger} \ket{0}=\ket{\downarrow \downarrow}$. Using the hard-core constraint $\hat{s}^\dagger \hat{s}+\hat{t}^\dagger_\alpha \hat{t}_\alpha=1$ 
we obtain
\begin{align}
\mathcal{H}&=-\sum_{i,\sigma} \mu_{\sigma}\hat{n}_{i\sigma} +V\sum_{\langle i,j\rangle}
\sum_{\sigma_1,\sigma_2} \sigma_{1} \sigma_{2} \medspace \hat{n}_{i\sigma_1} \hat{n}_{j\sigma_2} \nn \\ 
&+t \sum_{\langle i,j\rangle} \left[ (\hat{t}_{i-}^{\dagger}-\hat{t}_{i+})(\hat{t}_{j-}-\hat{t}_{j+}^{\dagger}) + \textrm{h.c.}\right] 
\end{align}
with $\hat{n}_{i\sigma}=\hat{t}^{\dagger}_{i\sigma}\hat{t}_{i\sigma}$, $t=(J_1-J_2)/2$, $V=(J_1+J_2)/2$, and 
$\mu_{\sigma}=-(J_0-D)+\sigma h$. The inclusion of an anisotropy $D>0$ is not crucial for our analysis but simplifies matters 
by allowing us to project out the $t_0$ triplet which is energetically unfavorable. 

The different Mott insulating states can be easily found in the atomic limit $t\to0$.  
For $|h|<J_0-D$, the occupation numbers of both triplets are zero, corresponding to a gapped non-magnetic state. 
For sufficiently strong fields the magnetization is fully saturated with exactly one triplon on every site, $m_+=1$ or 
$m_-=1$,  depending on the sign of $h$. We label these states as $m=\pm1$ Mott insulators, respectively. Between the 
non-magnetic and fully polarized states, repulsive interactions between triplons on neighboring sites stabilize checkerboard 
order where every second site remains empty ($m=\pm 1/2$). Dimer couplings beyond nearest neighbors  
lead to additional incommensurate states with filling fractions that crucially depend on the lattice geometry. 
In all Mott insulating phases, the magnetization does not change as a function of field, giving rise to magnetization plateaus. 

\begin{figure}[t!]
\begin{center}
\includegraphics[width=\linewidth]{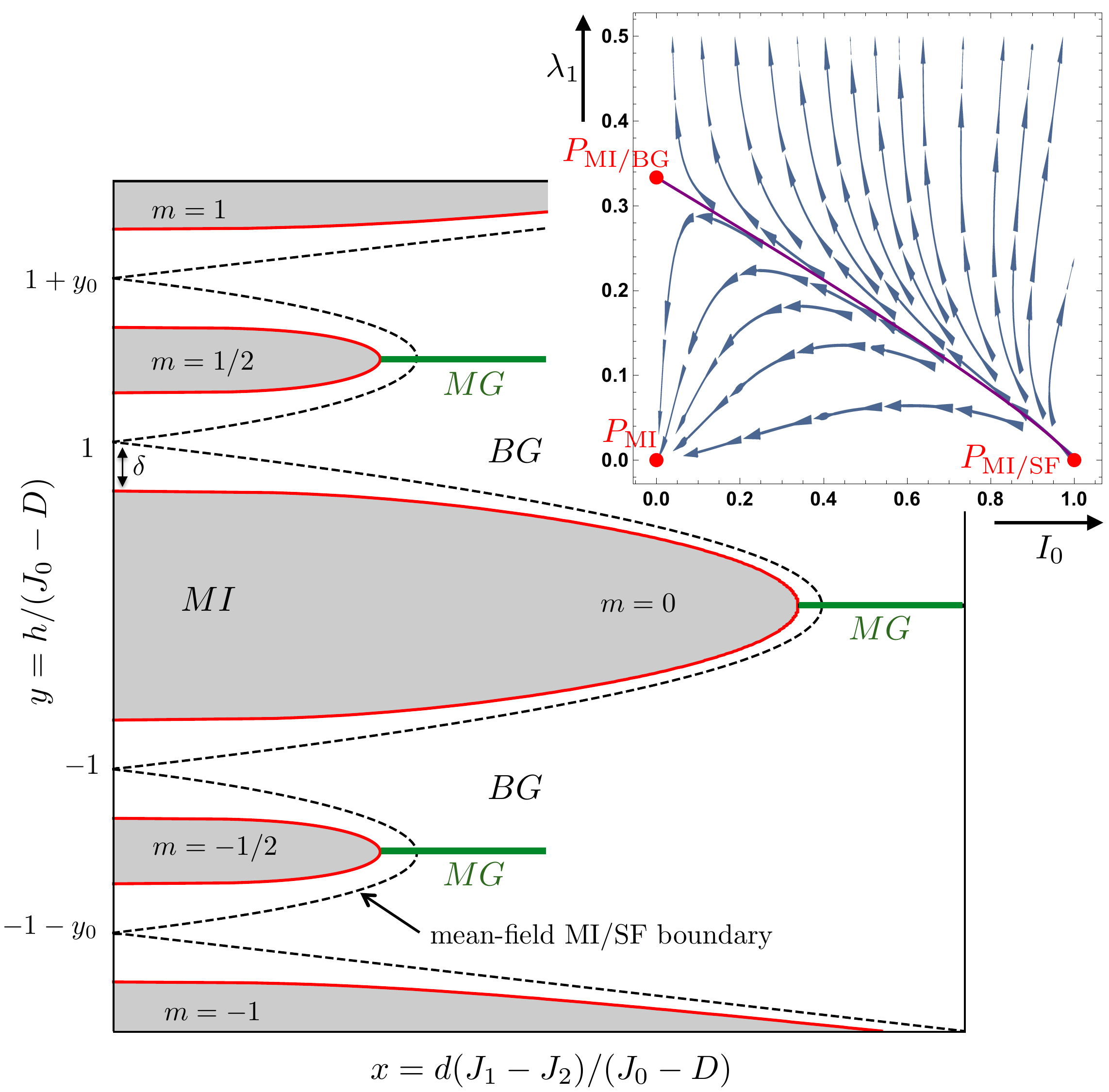}
\caption{(Color online) Phase diagram as a function of dimer coupling $x$ and magnetic field $y$ for $y_0=d(J_1+J_2)/(J_0-D)=1$. 
Dashed lines show the MI/SF mean-field transitions of the clean system.  Disorder leads to the formation of an 
incompressible Bose glass (BG) between the MI and the SF, which turns into an incompressible Mott glass (MG) at the tips 
of the Mott lobes. Solid red lines are the MI/BG phase boundaries obtained from an RG analysis in $d=3$ for a disorder 
strength of $\delta=\Delta/(J_0-D)=0.3$. 
Inset: RG flow of the inverse mean $I_0=1/(1+\overline{r})$, and  relative variance 
$\lambda_1\sim (\overline{r^{2}}-\overline{r}^2)/\overline{r}^2$ of the  random-mass distribution. 
There are three fixed points: a stable MI fixed point at $(I_0,\lambda_1)=(0,0)$, a MI/SF transition at $(1,0)$ and a 
MI/BG transition at $(0,d/9)$. The BG/SF transition is not accessible in our strong-coupling approach.}
\label{fig.2}
\end{center}
\end{figure}

For large enough $t$, the system becomes a superfluid, corresponding to a canted $XY$ 
antiferromagnet. The phase boundaries are obtained from a strong-coupling expansion around the atomic limit.  
Performing a Hubbard-Stratonovich decoupling of the hopping term, we obtain a dual continuum action for
the superfluid order parameter $\psi(\br,\tau)$ in space and imaginary time  \cite{Kruger+09,Kruger+11,Thomson+14}, 
\begin{equation}
\mathcal{S}_0  = \int_{\bk\omega}  K(\bk,\omega) |\psi(\bk,\omega)|^2 +u\int_{\br\tau}|\psi{(\br,\tau)}|^{4}.
\end{equation}
with $K(\bk,\omega)=(k^2- i \gamma_1\omega + \gamma_2\omega^2 +r)$ in the momentum and frequency domain. 
The mass $r$ and the interaction vertex $u$ are related to the local single- and two-particle Green functions for the 
different Mott insulating states respectively, e.g. $r=r_m=1+2td G_m(\omega=0)$ \cite{SM}. The MI/SF 
mean-field phase boundaries  are obtained by $r=0$ and shown as dashed lines in Fig.~\ref{fig.2} as a function of 
dimensionless dimer coupling $x$ and magnetic field $y$. The phase diagram has the familiar lobe structure of Bose-Hubbard models
and is symmetric around $h=0$. It has been suggested \cite{Batrouni+00,Yamamoto+12,Yamamoto+13} that the tips of the 
fractionally-filled lobes may exhibit first-order or supersolid behavior, however such a question is beyond the reach of the present analysis.

The coefficients of the frequency terms are given by derivatives of the mass coefficient with respect to the magnetic field, 
$\gamma_{1}=-\partial r/\partial y$ and $\gamma_{2}=-\frac12 \partial^2 r/\partial y^2$ \cite{note2}. At the tips of the Mott lobes, the slope 
$\gamma_1$ vanishes and the field theory becomes relativistic, reflecting the particle-hole symmetry at these points. 
Previous works \cite{Kruger+09,Kruger+11,Thomson+14} studying other aspects of the model have largely neglected the frequency terms, 
however we retain them here to study how the behavior changes near the tips. These frequency terms will turn out to be crucial to correctly 
describe the thermodynamics, a feature overlooked by the earlier studies.

The key question we are trying to answer in the present work is whether the glassy phase formed in disordered Bose-Hubbard models is 
always a compressible Bose glass \cite{Fisher+89}, or if the more elusive incompressible Mott glass may exist at the high-symmetry tips 
of the Mott lobes \cite{Roscilde+07,Sengupta+07,Ma+14,Wang+15}. First seen in 1d fermion systems \cite{Orignac+99,Giamarchi+01}, 
the Mott glass has also been predicted to exist in the  $O(2)$ quantum rotor model \cite{Altman+04,Altman+08,Altman+10,Iyer+12}, which maps 
to the Bose-Hubbard model at commensurate fillings, and has been experimentally observed in the disordered quantum 
antiferromagnet DTN \cite{Yu+12}.
 
We focus on random mass disorder such that $\mu_{i,\sigma}= \mu_{\sigma}+\varepsilon_i$. This can come from disorder in the intra-dimer 
coupling $J_0$, the anisotropy $D$ or the applied field $h$. In the following, we assume that the disorder has a symmetric box distribution 
of width $2\Delta$ and is uncorrelated between different sites. For sufficiently bounded disorder the phase diagram retains Mott insulating regions,
e.g. a central Mott lobe is present for $\delta=\Delta/(J_0-D)<1$. The Hubbard-Stratonovich transformation is performed in the same way, 
leading to disorder in all coefficients of the dual action. We use the replica trick \cite{Mezard+91} to obtain the disorder averaged free energy. 
The replicated action
\begin{equation}
\mathcal{S}  = \sum_{\alpha=1}^n\overline{\mathcal{S}_0[\psi^*_\alpha, \psi_\alpha]} - 
\frac{g}{2} \sum_{\alpha\beta}\int_{\br\tau\tau'}\hspace{-10pt} |\psi_\alpha(\br,\tau)|^2 |\psi_\beta(\br,\tau')|^2
\end{equation}
consists of two parts. The first contribution is simply $n$ copies of the original action with disorder averaged coefficients $\overline{\gamma}_1$, 
$\overline{\gamma}_2$, $\overline{r}$, and $\overline{u}$. The second term is the disorder vertex, which is non-diagonal in replica space 
$\alpha, \beta$ and imaginary time and proportional to the \emph{variance} of the random-mass distribution 
$g=(\overline{r^{2}}-\overline{r}^2)$.

To determine the phase diagram in the presence of weak disorder, we use a momentum-shell RG approach. 
As in previous work \cite{Kruger+09,Kruger+11,Thomson+14}, we make the change of 
variables $I_0=1/(1+\overline{r})$, and introduce the relative disorder variance $\lambda=I_{0}^{2} g$ to  distinguish between the Mott insulating and 
glassy phases. In all but the superfluid phase, $I_0$ flows to zero, reflecting the short-ranged superfluid correlations. The relative variance $\lambda$ 
compares the shift of the random mass distribution with its spread. In the Mott insulator, the distribution 
shifts faster than it spreads and $\lambda$ renormalizes to zero. If the spread is faster than the shift, the tail of the distribution pushes through zero, 
indicating a glassy phase where the physics is dominated by rare superfluid regions. Taking into account the 1-step RSB in this model \cite{Thomson+14}, 
the RG equations are:
\begin{subequations}
\label{eq.RG}
\begin{eqnarray}
I_0'(\ell) & = & ( 3/2 \lambda_1-2 )I_{0}+2I_{0}^{2},\\
\lambda_1'(\ell) & = & (4I_{0}-d)\lambda_1+9\lambda_1^{2},\\
 \overline{\gamma}_1'(\ell) & = & (2-z+\lambda_1)\overline{\gamma}_1, \\
 \overline{\gamma}_2'(\ell) & = & (2-2 z+\lambda_1)\overline{\gamma}_{2}+\lambda_1 I_{0} \overline{\gamma}_{1}^{2},
\end{eqnarray}
\end{subequations}
where $z$ is the dynamical critical exponent and $\lambda_0\equiv 0$ and $\lambda_1=I_0^2 g_1$ denote the step heights of the Parisi disorder 
function \cite{SM,Mezard+91}. We neglect the $\overline{u}$ vertex since it is irrelevant away from the tips 
($\overline{\gamma}_1\neq 0$). The RG flow in the $I_0$-$\lambda_1$ plane is shown in the inset in Fig.~\ref{fig.2}. 
The MI/SF fixed point is unstable against disorder, confirming that even for infinitesimal disorder, the transition from the Mott insulator is into a disordered 
insulating state and not into a superfluid \cite{Pollet+09,Gurarie+09}. As we show later, the disordered state is a compressible Bose glass, except for the tips 
where we see strong indications for an incompressible Mott glass state. 

The relative variance $\lambda_1$ diverges in the Bose glass. We stop integration at a scale $\ell^*$ where $\lambda_1(\ell^*)=1$ and our RG 
becomes invalid. This scale can be identified with a correlation length $\xi\sim e^{\ell^*}$ which corresponds to the typical separation of superfluid regions. Note that 
$\xi$ is not the superfluid correlation length, which remains finite at the MI/BG transition. Linearizing near 
$P_\textrm{MI/BG}$, we find the correlation length diverges as $\xi \sim (x-x_c)^{-1/d}$.

From the dependence of $I_0$ and $\lambda_1$ on the microscopic parameters, we can determine the MI/BG phase boundary as a function 
of the dimensionless dimer coupling $x$ and magnetic field $y$ for a given disorder strength $\delta$ 
(Fig.~\ref{fig.2}). As in the conventional Bose-Hubbard model, the Mott insulating regions shrink when disorder is added, and a glassy phase intervenes between 
the Mott insulating and the superfluid regions. In the original magnetic language, this corresponds to a spin glass phase.

To determine the nature of the glassy phase, we calculate the compressibility. For a single component boson system, the local compressibility is defined as the 
derivative of the local density with respect to the chemical potential and related to the particle-number fluctuations, 
$\kappa_i=\partial\langle \hat{n}_{i} \rangle/\partial \mu=\beta \sum_j (\langle \hat{n}_{i}\hat{n}_{j} \rangle -\langle \hat{n}_{j} \rangle \langle \hat{n}_{j} \rangle )$. 
This is easily generalized to the two-triplon case, where the compressibility is defined as the derivative of the local magnetization with respect to magnetic field. Performing a disorder 
average we can calculate $\kappa=-\partial^{2}\overline{\mathcal{F}}/\partial y^2$ directly from the replica field theory, yielding \cite{SM}
\begin{equation}
\label{eq.comp}
\kappa = \frac{2 S_d}{(2 \pi)^d} \frac{I_0^2\overline{\gamma}_2^2 \lambda_1}{(I_0^2 \overline{\gamma}_1^{2}+4I_0 \overline{\gamma}_2)^{3/2}},
\end{equation}
where $S_d$ is the surface of a $d$ dimensional unit sphere. It is crucial here to include the effects of 1-step RSB, otherwise $\kappa$ vanishes identically as a consequence 
of the frequency structure that is inherent to all Bose-Hubbard models \cite{SM,note2}. 
This remarkable result reveals the surprising physical importance of RSB in this system. 
\begin{figure}
\begin{center}
\includegraphics[width=\linewidth]{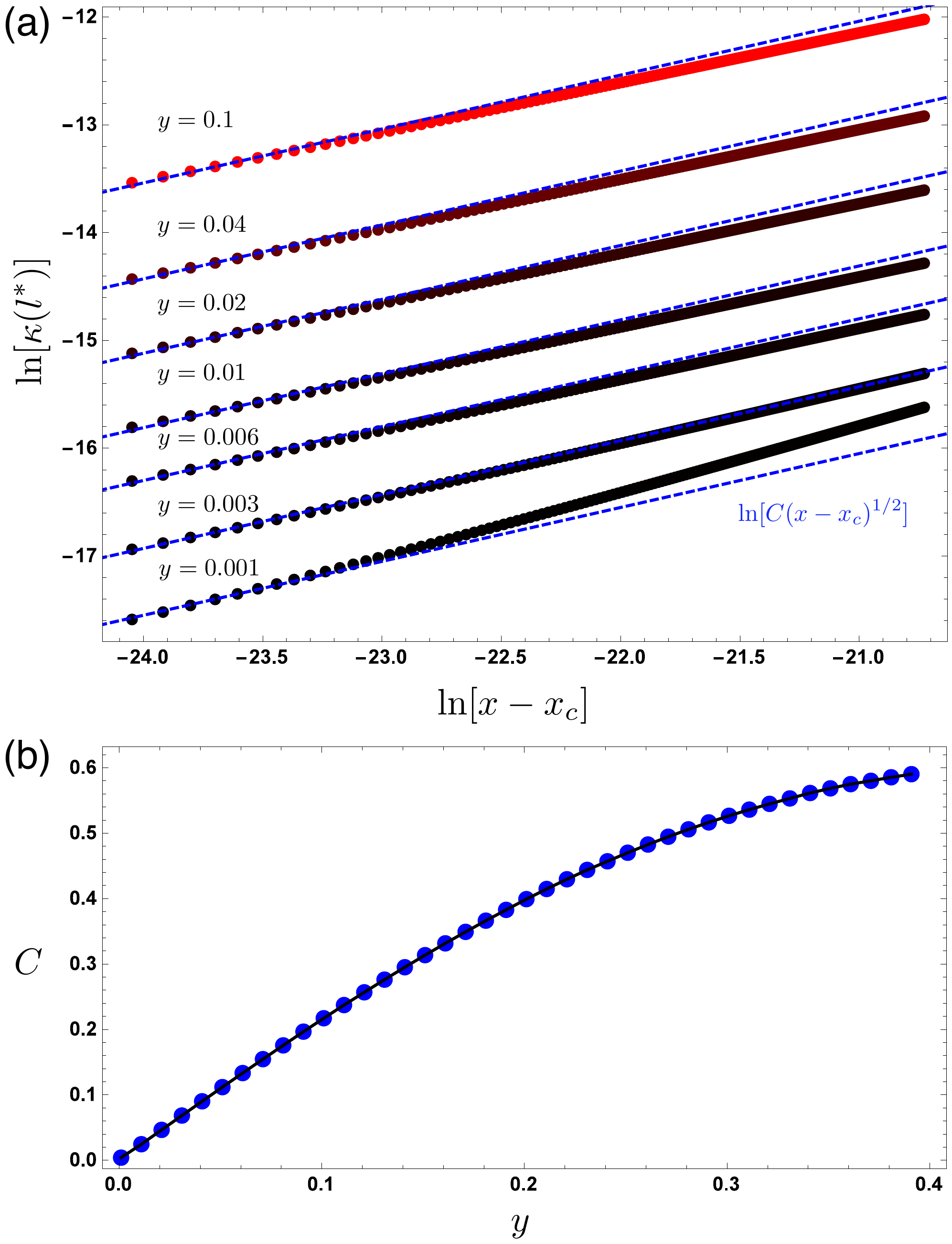}
\caption{(Color online) (a) Power-law behavior of the compressibility $\kappa(l^{*})\simeq C(x-x_c)^{1/2}$ ($d=3$) of the Bose glass close to the transition to the Mott insulator. 
The blue dashed lines are guides to the eye with gradient $1/2$. 
Near $y=0$ there is an anomalous suppression of the range of universal behavior.
(b) $C$ versus $y$, showing the vanishing compressibility close to the tip of the central lobe at $y=0$.}
\label{fig.3}
\end{center}
\end{figure}
Linearising around $P_\textrm{MI/BG}$ we find that at the transition to the Mott insulator the compressibility vanishes as
\begin{equation}
\kappa =\kappa(\ell^*)=C (x-x_c)^{\frac{2}{d}-\frac{1}{6}},
\label{eq.crit}
\end{equation}
where $C\sim |\overline{\gamma}_1|$. Away from the tip, the disordered state adjacent to the Mott insulator is a compressible Bose glass. Approaching a particle-hole 
symmetric Mott lobe tip, the coefficient $C$ vanishes, suggesting a change of universality. 
The analytical result (\ref{eq.crit}) is confirmed by a numerical calculation of $\kappa(\ell^*)$ from the full RG equations (\ref{eq.RG}) for the central Mott lobe in 
$d=3$ with $\delta=0.3$ (see Fig.~\ref{fig.3}a).  The coefficient $C$ has a strong dependence on the field and vanishes linearly as $y\to 0$ (Fig.~\ref{fig.3}b). 
Very close to the tip at $y=0$, there is a strong suppression of the range of universal behavior, but no indication of a crossover to a different universality class.
This highlights the singular nature of this point and strongly indicates the existence of an incompressible Mott glass state.


Although we have focused on dimerized quantum antiferromagnets using the hard-core boson formalism, our prediction of a Mott glass is valid across a wide 
range of systems, including conventional Bose-Hubbard and Jaynes-Cummings Hubbard \cite{Angelakis+07} models. Any such Bose-Hubbard-like model  may be treated 
using the methods outlined here, with the microscopic differences appearing only in the UV-scale starting values of 
the flow parameters. This prediction lends weight to previous numerical quantum Monte-Carlo work \cite{Sengupta+07,Wang+15}.
It may also explain the controversy over the existence of a direct MI/SF transition at the tips of the Mott lobes in the disordered Bose-Hubbard model: 
previous works which used compressibility as the criterion for the onset of a glassy phase will necessarily 
have missed the transition between the Mott insulator and the Mott glass.  

The breakdown of self-averaging \cite{Kruger+11,Hegg+13,Zuniga+15}, the importance of replica symmetry breaking \cite{Thomson+14}, and the connection with spin-glass 
phenomena strongly suggest that the formation of  Bose and Mott glass phases is not simply a weak localization phenomenon. It would be interesting to review the nature 
of these phases in the context of many-body localization \cite{Nandkishore+15} and the related entanglement entropy scaling \cite{Kjall+14}.

The equivalence between Bose-Hubbard models and dimerized quantum antiferromagnets allows for multiple complementary experiments to verify our theoretical 
predictions. Previous measurements on disordered ultracold atomic gases have inferred the presence of the Bose glass from macroscopic measurements 
\cite{Lye+05,Fallani+07,Meldgin+15}, however the boson number fluctuations associated with the local compressibility are also now within reach of quantum 
gas microscope systems \cite{Bakr+09,Sherson+10}, potentially allowing for direct imaging of the glassy phases. Thermodynamic measurements of the 
magnetic susceptibility and the specific heat  of dimerized quantum antiferromagnets can provide clear signatures of Bose and Mott glass phases \cite{Yu+12}. 
We also expect to see characteristic differences in the glassy dynamics, which could be studied with $\mu$SR, 
as well as in the magnetic excitation spectra. Dimer systems exhibiting geometric frustration are a particularly intriguing theoretical 
problem for further study, as are additional types of disorder such as non-magnetic impurities \cite{Samulon+11}.

\acknowledgements
The authors would like to thank B. Braunecker, G. D. Bruce, C. Hooley, J. Keeling, and C. Pedder for valuable discussions. SJT acknowledges financial support from the Scottish CM-CDT.

\appendix

\clearpage 

\section{Supplementary Material}

In this Supplementary Material to our Letter we go through the technical details of our calculation. We show in detail  the construction of the strong coupling field theory and 
the calculation of the compressibility. In particular, we wish to stress the importance of including replica symmetry breaking effects to correctly describe the disordered phases.

\section{Mapping to Hard-Core Bosons}

The spin operators of each dimer may be written
 in terms of the singlet and triplet bond operators as: 
\begin{align}
\hat{S}_{1,2}^{\alpha}&=\frac12 \left[ \pm \hat{s}^{\dagger} \hat{t}_{\alpha} \pm \hat{t}_\alpha^\dagger \hat{s}-
i \epsilon_{\alpha \beta \gamma}\hat{t}_{\beta}^{\dagger} \hat{t}_{\gamma} \right],
\end{align}
where $\hat{s}^{\dagger}\ket{0} = \frac{1}{\sqrt{2}}(\ket{\uparrow \downarrow}-\ket{\downarrow \uparrow})$, 
$\hat{t}_{x}^{\dagger}\ket{0}=-\frac{1}{\sqrt{2}}(\ket{\uparrow \uparrow}-\ket{\downarrow \downarrow})$, 
$\hat{t}_{y}^{\dagger}\ket{0} =\frac{i}{\sqrt{2}}(\ket{\uparrow \uparrow}+\ket{\downarrow \downarrow})$, and 
$\hat{t}_{z}^{\dagger}\ket{0} =\frac{1}{\sqrt{2}}(\ket{\uparrow \downarrow}+\ket{\downarrow \uparrow})$. 
These operators are bosonic and subject to the hard-core constraint 
$\hat{s}^{\dagger}\hat{s}+\hat{t}^{\dagger}_{\alpha}\hat{t}_{\alpha}=1$. The local dimer Hamiltonian is diagonal 
in the standard triplet basis, $\hat{t}_{+}^{\dagger}\ket{0}=\ket{\uparrow \uparrow}=\ket{1,1}$, 
$\hat{t}_{-}^{\dagger} \ket{0}=\ket{\downarrow \downarrow}=\ket{1,-1}$, and $\hat{t}_{0}^{\dagger}\ket{0}=\hat{t}_{z}^{\dagger} \ket{0}=\ket{1,0}$. The Hamiltonian (Eq. 2) in our Letter is obtained by a change of variables to this basis.  Note that we have also used the hard-core constraint to eliminate the singlet 
operators and to rewrite on-site products of number operators, $\hat{n}^{2}_{i\sigma}=\hat{n}_{i\sigma}$ and $\hat{n}_{i-}\hat{n}_{i+}=0$.

\section{Strong Coupling Field Theory}
We begin from the bosonic Hamiltonian shown in our Letter (Eq. 3) and use it to write the partition function at $T=0$ as an imaginary time path integral over bosonic coherent states, 
$\mathcal{Z}=\int\mathcal{D}[\overline{\phi},\phi]e^{-(\mathcal{S}_0+\mathcal{S}_t)}$, where the local and kinetic contributions to the action are given by
\begin{subequations}
\begin{eqnarray}
\mathcal{S}_0 [\overline{\phi},\phi]& = & \int_{0}^{\infty} \mathrm{d} \tau \Big\{ \sum_{i,\sigma=\pm} (\overline{\phi}_{i \sigma}(\tau)(\partial_{\tau}-\mu_{\sigma})\phi_{i \sigma}(\tau) \nn\\
& &  +V \sum_{\langle i,j \rangle} \sum_{\sigma_1,\sigma_2}\sigma_1 \sigma_2 |\phi_{i \sigma_1}(\tau)|^{2}|\phi_{j \sigma_2}(\tau)|^{2} \Big\}, \quad \\
\mathcal{S}_t[\overline{\phi},\phi] & = & \int_{0}^{\infty} \mathrm{d} \tau\sum_{i,j}T_{ij} \left[ \overline{\phi}_{i-}(\tau)-\phi_{i+}(\tau)  \right] \nn\\
& & \hspace{40pt}\times \left[ \phi_{j-}(\tau)-\overline{\phi}_{j+}(\tau)  \right].
\end{eqnarray}
\end{subequations}

The hopping matrix elements $T_{ij}$ are equal to $t$ if $i,j$ are nearest-neighbor sites and zero otherwise. To facilitate the strong-coupling expansion around the local 
limit ($t\to 0$), we decouple the kinetic term by introducing auxiliary Hubbard-Stratonovich fields $\overline{\psi}_i(\tau)$, $\psi_i(\tau)$,
\begin{subequations}
\begin{eqnarray}
e^{-\mathcal{S}_t[\overline{\phi},\phi]} & = & \int\mathcal{D}[\overline{\psi},\psi] e^{-( \tilde{\mathcal{S}}_t[\overline{\psi},\psi]+\mathcal{S}'_{\phi\psi}  )},\\
\tilde{\mathcal{S}}_t[\overline{\psi},\psi] & = & \int_\tau \sum_{ij}T^{-1}_{ij} \overline{\psi}_{i}(\tau)  \psi_{j}(\tau),\\
\mathcal{S}'_{\phi\psi}& = & \int_\tau \sum_{i}   [\overline{\phi}_{i-}(\tau)-\phi_{i+}(\tau)]\psi_{i}(\tau)+\textrm{h.c.}\quad
\end{eqnarray}
\end{subequations}

Because of the structure of the initial action, we only require a single Hubbard-Stratonovich field to perform the decoupling, despite the presence of two bosonic fields in the 
initial action. We can now trace out the original fields to obtain the partition function in terms of the new fields, 
$\mathcal{Z}= \int \mathcal{D}[\overline{\psi},\psi]e^{-\tilde{\mathcal{S}}[\overline{\psi},\psi]}$. The resulting dual 
action is given by $\tilde{\mathcal{S}}[\overline{\psi},\psi] =  \tilde{\mathcal{S}}_t[\overline{\psi},\psi]-\delta\tilde{\mathcal{S}}[\overline{\psi},\psi]$ with
\begin{eqnarray}
\delta\tilde{\mathcal{S}} & = &  \mathrm{ln} \left\langle \mathcal{T} \exp\Big\{ \int_\tau \sum_{i} [\overline{\psi}_{i}(\tau)\hat{b}_i(\tau) +\hat{b}_i^\dagger(\tau)\psi_{i}(\tau) ] \Big\} \right\rangle_0\quad
\end{eqnarray}
where $\mathcal{T}$ denotes the time-ordering operator and the average is taken with respect to the local Hamiltonian, 
\begin{equation}
\mathcal{H}_0 =  -\sum_{i,\sigma=\pm}\mu_\sigma \hat{n}_{i\sigma} +V \sum_{\langle i,j \rangle} \sum_{\sigma_1,\sigma_2}\sigma_1 \sigma_2  \hat{n}_{i\sigma_1} \hat{n}_{j\sigma_2},
\end{equation}
$\hat{n}_{i\sigma}=\hat{t}_{i\sigma}^\dagger\hat{t}_{i\sigma}$. For brevity, we have defined $\hat{b}_i(\tau)=\hat{t}_{i-}(\tau)-\hat{t}_{i+}^{\dagger}(\tau)$ and 
$\hat{t}_{i\sigma}(\tau) = e^{\mathcal{H}_0\tau}\hat{t}_{i\sigma}e^{-\mathcal{H}_0\tau}$.
In principle, we can expand $\delta\tilde{\mathcal{S}}$ to 
any desired order in the fields $\psi$. However, since $\mathcal{H}_0$ contains interactions, we cannot use Wick's theorem to calculate expectation values of products of the original bosonic operators. 
The expansion only contains powers of $|\psi|^2$ since $\mathcal{H}_0$ preserves the local particle number.  Expanding $\delta\tilde{\mathcal{S}}$ up to quartic order, we obtain
\begin{equation}
\delta\tilde{\mathcal{S}}  =   \sum_i \int_\omega G(\omega)  |\psi_i(\omega)|^2+U\sum_i\int_\tau |\psi_i(\tau)|^4,
\end{equation}
where $G(\omega)$ is the Fourier transform of the single particle Green function 
\begin{equation}
G(\omega) = -\int_{-\infty}^{\infty}  \mathrm{d} \tau e^{i\omega \tau} \langle \mathcal{T} \hat{b}_i(0) \hat{b}_i^\dagger(\tau) \rangle_0.
\end{equation}
It is straightforward to calculate $G_m(\omega)$ for the different MI states,
\begin{subequations}
\begin{eqnarray}
G_{\pm1}(\omega) & = & -\frac{1}{\mu_{\pm}-zV+i \omega},  \\
G_{\pm1/2}(\omega) &= & -\frac{1}{\mu_{\pm}+i \omega}-\frac{1}{-\mu_{\pm}+zV-i \omega}\nn \\ 
& & +\frac{1}{\mu_{\mp}+zV-i \omega},  \\
G_{0}(\omega) & = & \frac{1}{\mu_+ +  i \omega}+\frac{1}{\mu_- -  i \omega}.
\end{eqnarray}
\end{subequations}

The coefficient $U$ of the quartic vertex is given by the connected parts of the two-particle Green functions. Unlike in the conventional Bose-Hubbard model, the calculation here 
is relatively simple because of the hard-core constraint. In each of the regions, $U_m$ is given by 
\begin{subequations}
\begin{eqnarray}
U_{\pm 1} & = &  \frac{1}{(\mu_{\pm}-2dV)^3}\nn\\
& & -\frac12 \frac{1}{(\mu_{\pm}-2dV)^{2} (\mu_{\pm}-\mu_{\mp}-4dV)},\\
U_{\pm1/2} & = &  \frac{1}{\mu_{\pm}^3}+\frac{1}{(2dV-\mu_{\pm})^3}-\frac{1}{(\mu_{\mp}+2dV)^{3}} \nn \\ 
& &  -\frac{\mu_{\pm}+\mu_{\mp}}{(\mu_{\pm}-2dV)^{2}(\mu_{\mp}+2dV)^{2}}\nn\\
& & -\frac12 \frac{1}{\mu_{\pm}^{2}(\mu_{\pm}-\mu_{\mp})},\\
U_{0} & = &  -\frac{1}{\mu_+^{3}}-\frac{1}{\mu_-^{3}}-\frac{\mu_+ + \mu_-}{\mu_{+}^{2}\mu_{-}^{2}}. 
\end{eqnarray}
\end{subequations}

Analyzing the above expression we find that $U$ is always positive. Hence there is no indication of a first-order instability of the $m=\pm1/2$ checkerboard lobes at this level.
The hopping term $\tilde{\mathcal{S}}_t$ can be treated  in the long-wavelength limit. In this case it is possible to invert the original hopping matrix and we 
obtain
\begin{equation}
\tilde{\mathcal{S}}_t \approx \int_{\bk\omega} \left( \frac{1}{2dt}+\frac{1}{4d^2t}k^{2} \right)|\psi(\bk,\omega)|^2.
\end{equation}

Combining $\tilde{\mathcal{S}}_t$ and the quadratic contribution to $\delta\tilde{\mathcal{S}}$ and expanding $G(\omega)$ for small frequencies we obtain 
\begin{equation}
\tilde{\mathcal{S}}_2 =  \int_{\bk\omega} \left(\frac{1}{4d^2t} k^2 + R -iK_1\omega+K_2\omega^2   \right) |\psi(\bk,\omega)|^2
\end{equation}
with $R=1/2dt + G(\omega=0)$. Inspecting $G(\omega)$ and remembering that $\mu_\sigma=-(J_0-D)+\sigma h$, we see that $G$ is a function 
of $h+i\omega$. From this, it immediately follows that $K_1=-\partial R/\partial h$ and $K_2=-\frac12 \partial^2 R/\partial h^2$. 

As a last step, we rescale the action to dimensionless units, using $\omega\to V\omega$, $\bk\to\sqrt{2d}\bk$, and $|\psi|^2\to \sqrt{2d}^{2-d}t/V\cdot|\psi|^2$.
This leads to the dimensionless strong-coupling action given in the Letter with coefficients $r=2dt R$, $\gamma_1=2dt VK_1$, $\gamma_2=2dtV^2 K_2$, and
$u=\sqrt{2d}^{4-3d}t^2 U/V^3$.

In terms of 
the dimensionless dimer coupling (hopping) $x=d(J_1-J_2)/(J_0-D)$ and magnetic field (chemical potential) 
$y=h/(J_0-D)$ the mass coefficients on the different regions are 
\begin{subequations}
\begin{eqnarray}
r_0 & = & 1-x[(1+y)^{-1}+(1-y)^{-1}],\\
r_{\pm 1/2} & = & 1-x[(1+y_0\mp y)^{-1}-(1\mp y)^{-1}\nn\\
& & +(1-y_0\pm y)^{-1}],\\
r_{\pm 1} & = & 1+x(1+y_0\mp y)^{-1}, 
\end{eqnarray}
\end{subequations}
where $y_0=d(J_1+J_2)/(J_0-D)$.

\section{Replica disorder average}

In the presence of on-site disorder $\epsilon_i$, which randomly shifts the local chemical potentials, $\mu_{i\sigma}=\mu_\sigma+\epsilon_i$, the Hubbard-Stratonovich 
transformation can be performed in exactly 
the same way. This results in a dual strong-coupling action on a lattice with disorder in all the coefficients, e.g. the mass on site $i$ is given by
\begin{equation}
r_i = r(\mu_\sigma+\epsilon_i,V,t),
\end{equation}
where $r(\mu_\sigma,V,t)$ is the the mass expression we have derived for the clean, homogeneous system. As discussed in the Letter, we use the replica trick $\overline{\mathcal{F}}=-T \overline{\ln\mathcal{Z}}=-T \lim_{n \rightarrow 0} \frac{1}{n}(\overline{\mathcal{Z}^{n}}-1)$ to 
average the free energy over the disorder. Here we provide some additional details. 
The effective replica action $\mathcal{S}_\textrm{eff}$ is obtained by taking $n$ copies of the system (labelled by $\alpha$) and performing the disorder average
\begin{eqnarray}
\overline{\mathcal{Z}^n} & = &  \int\mathcal{D}[\overline{\psi}_\alpha,\psi_\alpha] \overline{e^{-\sum_\alpha \mathcal{S}[\overline{\psi}_\alpha,\psi_\alpha] }} \nn\\
& & =: \int\mathcal{D}[\overline{\psi}_\alpha,\psi_\alpha] e^{- \mathcal{S}_\textrm{eff}[\{\overline{\psi}_\alpha,\psi_\alpha\}]}.
\end{eqnarray}

Expanding the exponential, performing the disorder average, and re-exponentiating, we obtain the general expression  
\begin{equation}
\mathcal{S}_\textrm{eff}=\sum_\alpha \overline{\mathcal{S}_\alpha}-\frac12\sum_{\alpha\beta}\left(\overline{ \mathcal{S}_\alpha \mathcal{S}_\beta} -
\overline{ \mathcal{S}_\alpha}\phantom{\cdot} \overline{ \mathcal{S}_\beta}   \right),
\label{eq.rep}
\end{equation}
where $\mathcal{S}_\alpha=\mathcal{S}[\overline{\psi}_\alpha,\psi_\alpha]$, for brevity. This expression is exact up to quartic order in the fields. The first term simply corresponds
to $n$ copies with disorder-averaged coffiecients $\overline{r}$, $\overline{\gamma}_1$, $\overline{\gamma}_2$, $\overline{u}$. For a box distribution $p(\epsilon)$ all the averages
can be performed analytically. E.g, for the central Mott lobe around zero field we obtain the average mass
\begin{equation}
\overline{r}_{m=0} = 1-\frac{x}{2\delta}\ln\left[\frac{(1+\delta)^2-y^2}{(1-\delta)^2-y^2}    \right],
\end{equation}
with $x$ the dimensionless hopping, $y$ the dimensionless field, and $\delta$ the dimensionless width of the box distribution, as defined in the Letter. Since the field derivatives
commute with the disorder average, the relations $\overline{\gamma}_1=-\partial \overline{r}/\partial y$ and $\overline{\gamma}_2=-\frac12 \partial^2 \overline{r}/\partial y^2$
still hold. It is important to include the moments of $p(\epsilon)$ to \emph{all} orders in order to account for the boundedness of the distribution. This is crucial for the presence of Mott 
insulating states in the disordered system.

The second term in Eq.~(\ref{eq.rep}) contains various quartic disorder vertices. We neglect all the vertices with temporal gradients as they are irrelevant under the RG, and keep only the 
disorder vertex that arrises from the replication of the mass term. It is obvious from Eq.~(\ref{eq.rep}) that the coefficient is given by the variance $g=\overline{r^2}-\overline{r}^2$ of the 
random mass distribution. This can also be calculated analytically for the box distribution $p(\epsilon)$.

\section{Renormalization and Replica Symmetry Breaking}

We integrate out the highest energy modes, corresponding to momenta in the infinitesimal shell $\mathrm{e}^{-\mathrm{d}l} \leq |\bk|\leq1$. To regain resolution, we rescale 
 momenta $\bk \rightarrow \bk \mathrm{e}^{-\mathrm{d}l}$ and frequencies $\omega\rightarrow\omega \mathrm{e}^{z \mathrm{d}l}$. The fields are 
 rescaled such that the $k^2$ term remains constant. This leads to coupled RG equations for $\overline{r}$, $\overline{\gamma}_1$, $\overline{\gamma}_2$, $\overline{u}$, and $g$. 
 For simplicity, we neglect the $\overline{u}$ vertex since it is irrelevant under the RG and only leads to a small shift of the phase boundaries. 
 We allow for replica symmetry breaking (RSB) and replace $g$ by a matrix $g_{\alpha \beta}$ of general Parisi hierarchical form \cite{Mezard+91}. In previous work \cite{Thomson+14} 
we examined this model and showed the presence of 1-step RSB. Here we will show that RSB is responsible for the finite compressibility of the Bose glass. 

Taking the replica limit, the matrix $g_{\alpha \beta}$ becomes a monotonically increasing function $g(u)$ 
($u \in [0,1]$), which describes how the off-diagonal matrix elements of $g_{\alpha \beta}$ decrease as we move away from the diagonal with elements 
$g_{\alpha \alpha}=\tilde{g}$. As we have discussed in the Letter, the relative variance is the relevant variable to distinguish the Mott insulator and the Bose glass.  We therefore rescale 
the Parisi matrix by $I_0^2$, $[\tilde{\lambda},\lambda(u)]=I_0^2 [\tilde{g},g(u)]$. Note that $I_0 =1/(1+\overline{r})$ is the on-shell propagator in the zero-frequency limit and for large $\overline{r}$ 
(deep in the insulating phases) becomes the inverse mean of the random mass distribution.
At one-loop order the RG equations are given by \cite{Thomson+14}

\begin{subequations}
\begin{eqnarray}
\frac{\ud I_0}{\ud \ell} & = &   \left[-2+\tilde{\lambda}+\rho_0 \left(\tilde{\lambda}-\langle \lambda \rangle\right)\right]I_{0}+2I_{0}^{2},\\
\frac{\ud \overline{\gamma}_1}{\ud \ell} & = & (2-z+\tilde{\lambda})\overline{\gamma}_1, \\
\frac{\ud \overline{\gamma}_2}{\ud \ell} & = & (2-2 z+\tilde{\lambda})\overline{\gamma}_{2}+ \tilde{\lambda} I_{0}  \overline{\gamma}_{1}^{2}, \\
\frac{\ud \tilde{\lambda}}{\ud \ell} & = &  (4 I_0 -d)\tilde{\lambda} + 6 \tilde{\lambda}^{2}\nn \\ 
& & +2\rho_0 \left[\left(\tilde{\lambda}-\langle \lambda \rangle\right) \tilde{\lambda} +2\left(\tilde{\lambda}^2-\langle \lambda^{2} \rangle\right) \right],\\
\frac{\ud \lambda(u)}{\ud \ell} & = & (4 I_0 -d) \lambda (u) + 2 \lambda^2(u) +4 \tilde{\lambda}\lambda(u)+4\rho_0\nn \\
& &  \times\left[\frac52  \left(\tilde{\lambda}-\langle \lambda \rangle\right) \lambda(u)- \int_{0}^{u} \ud v \left[\lambda(u)-\lambda(v)\right]^{2} \right], \nn\\
\end{eqnarray}
\label{eqs.RG}
\end{subequations}
where $\langle \lambda^n \rangle=\int_{0}^{1}\lambda^n(u) \ud u$ and 
\begin{eqnarray}
\rho_0 & = & \int_\omega I_0^{-1}(1+r-i \overline{\gamma}_1 \omega+\overline{\gamma}_2 \omega^2)^{-1}\nn\\
&  = & \frac{1}{\sqrt{I_0^2\overline{\gamma}_1^2+4I_0\overline{\gamma}_2}}.
\end{eqnarray}

Analysis of the RG equations (\ref{eqs.RG}) shows that the system is unstable towards RSB and that the only stable solutions have 1-step RSB. To be more precise, $\lambda(u)$ is a step 
function with $\lambda(u<u_c)=\lambda_0=0$ and $\lambda(u>u_c)=\lambda_1=\tilde{\lambda}$ where the step position is $u_c=1/2\rho_0$ \cite{Thomson+14}. Specifying to 
this step function, we obtain the set of RG equations given in the Letter.

\section{Compressibility}

In this section we will provide details of the compressibility calculation and show that the finite compressibility of the Bose glass is a consequence of RSB. The compressibility 
is defined as
\begin{equation}
\kappa =- \frac{\partial^{2} \overline{\mathcal{F}}}{\partial y^2}=T \lim_{n \rightarrow 0} \frac{1}{n} \frac{\partial^{2} \overline{\mathcal{Z}^{n}}}{\partial y^2},
\end{equation}
where $y$ denotes the dimensionless magnetic field. Taking into account that all the coefficients in the quadratic replica action are functions of $y$ we obtain

\begin{eqnarray}
\kappa & = &  \lim_{n \rightarrow 0} \frac{1}{n} \left\{ 2 \overline{\gamma}_2 \sum_{\alpha} \int_{k,\omega} \langle |\psi_{\alpha}(k,\omega)|^{2} \rangle \right.\nn\\
& &+ \sum_{\alpha \beta} \int_{\bk\bk'\omega\omega'} (\overline{\gamma}_1+2i \omega \overline{\gamma}_2)(\overline{\gamma}_1+2i \omega' \overline{\gamma}_2) \nn\\
& &  \left.\phantom{ \int_{k,\omega}}\times \langle |\psi_{\alpha}(k,\omega)|^{2}|\psi_{\beta}(k',\omega')|^{2} \rangle\right\},
\end{eqnarray}
where the averages are taken with respect to the full action with interaction and disorder vertices $\overline{u}$ and $g_{\alpha\beta}$. Note that we have kept 
terms up to quadratic order in frequency as we did in the derivation of the strong coupling action. We compute the expectation values by expanding to linear order in the vertices, 
$\langle\ldots\rangle\approx \langle\ldots\rangle_0-\langle\ldots \mathcal{S}_u\rangle_0-\langle\ldots \mathcal{S}_g\rangle_0$, which results in three additive contributions
to the compressibility, $\kappa=\kappa_0+\kappa_u+\kappa_g$. Note that the replica limit $n\to 0$ ensures that only connected diagrams contribute. 

Let us first calculate the compressibility $\kappa_0$ for $\overline{u}=g_{\alpha\beta}=0$. Taking the expectation values with respect to the quadratic action,
 $\langle\ldots\rangle =\langle\ldots\rangle_0$, we immediately obtain
\begin{equation}
\kappa_0 = \left[2\overline{\gamma}_2 -\overline{\gamma}_{1}^{2} \frac{\partial}{\partial \overline{r}} + 4 \overline{\gamma}_1 \overline{\gamma}_2 \frac{\partial}{\partial \overline{\gamma}_1}+4\overline{\gamma}_2^2 \frac{\partial}{\partial \overline{\gamma}_2} \right] I_1,
\label{eq.kappa0}
\end{equation}
where $I_1$ denotes the integral of the correlation function over momentum and frequency, $I_1=\int_{\bk\omega} \left(k^2+\overline{r}-i \overline{\gamma}_1 \omega + \overline{\gamma}_2 \omega^2\right)^{-1}$. Since in the insulating phases $\overline{r}\to\infty$ under the RG, we can approximate the momentum integral for $|\bk|<1\ll \overline{r}$, and carry out the frequency 
integration to obtain
\begin{equation}
I_1=\frac{S_d}{(2\pi)^d}\frac{1}{\sqrt{\overline{\gamma}_1^2+4(1+\overline{r})\overline{\gamma}_2}}
\end{equation}
with $S_d$ the surface of the $d$-dimensional unit sphere. Taking the derivatives in Eq.~(\ref{eq.kappa0}), we find that $\kappa_0$ vanishes identically. The same is true for $\kappa_u$ since 
it is proportional to further derivatives of $\kappa_0$. This simply reflects that the gapped MI is incompressible. 

A finite compressibility has to be a consequence of the diverging disorder vertex, $\kappa=\kappa_g$. Starting from a general Parisi matrix $g_{\alpha\beta}$, we obtain
\begin{equation}
\kappa = \frac{S_d}{(2\pi)^2}\left(\tilde{\lambda}-\langle \lambda \rangle\right)\frac{4I_0^2 \overline{\gamma}_{2}^{2}}{(I_0^{2} \overline{\gamma}_{1}^2+I_0\overline{\gamma}_{2})^{2}},
\label{eq.kappag}
\end{equation}
where $[\tilde{\lambda},\lambda(u)]$ represents $\lambda_{\alpha\beta}=I_0^2 g_{\alpha\beta}$ after taking the replica limit, as we have explained above. 
This result is remarkable. It demonstrates that without RSB, $g_{\alpha\beta}\equiv g$ or $\lambda(u)\equiv \tilde{\lambda}$, the compressibility vanishes. This demonstrates 
that RSB is essential to correctly describe the physical properties of the Bose glass phase. 

For the stable 1-step RSB solution, we obtain $\tilde{\lambda}-\langle\lambda\rangle = u_c \lambda_1$ with $\lambda_1$ the step hight and 
$u_c=1/2 \rho_0=\frac12 \sqrt{I_0^{2} \overline{\gamma}_{1}^2+I_0\overline{\gamma}_{2}}$ the step position. Substituting into Eq.~(\ref{eq.kappag}) we arrive at the final expression 
provided in our Letter.

\section{Critical Scaling}

Using a scale-dependent dynamical exponent $z(\ell)=2+\lambda_1(\ell)$, we enforce that $\overline{\gamma}_1$ 
does not scale under the RG. In the vicinity of $P_\textrm{MI/BG}$ we obtain $\overline{\gamma}_2(\ell)=\frac25 c_0 \overline{\gamma}_1^{2}\exp[-(2-d/6)\ell]+c_1 \exp[-(2+d/9)\ell]$. 
If we are not exactly at the tip ($\overline{\gamma}_1\neq 0$), the first term will dominate for sufficiently large $\ell$ and $ \overline{\gamma}_2(l) \sim \overline{\gamma}_1^2 I_0(l)$. 
Substituting into Eq.~(6) in the main text and evaluating at the correlation length $\xi \sim e^{\ell^*} \sim (x-x_c)^{-1/d}$, we find that at the transition to the Mott insulator the compressibility vanishes as $\kappa =\kappa(\ell^*)=C (x-x_c)^{\frac{2}{d}-\frac{1}{6}}$ as stated in our Letter.


\begin{thebibliography}{63}
\bibitem[{ani()}]{anisotropy}
\bibinfo{note}{Note that single-ion anisotropy
  $\sim[(\hat{S}_{i1}^z)^2+(\hat{S}_{i2}^z)^2]$ is irrelevant in spin-1/2 dimer
  systems as it does not lift the degeneracy of the triplets. Our anisotropy
  $D$ is equivalent to a term $2D \hat{S}_{i1}^z \hat{S}_{i2}^z$ ($xxz$
  intra-dimer coupling) and acts like a single-ion anisotropy for the spin-1
  triplet excitations.}
\bibitem[{not()}]{note2}
\bibinfo{note}{The frequency dependence in the quadratic action of the dual
  field theory enters through the local Green function that is a function of
  $y+i\omega$, where $y$ is the dimensionless chemical potential or the
  equivalent magnetic field. From this it follows immediately that the
  frequency coefficients $\gamma_i$ are give by $i$th order derivatives of the
  mass coefficient $r$ with respect to $y$.}
\end{thebibliography}
\end{document}